\documentclass[a4paper,11pt]{article}
\usepackage{jheppub}
\usepackage{amsmath,amssymb,amsfonts,amsthm}
\usepackage{braket}
\usepackage{hyperref}
\title{Holographic Mutual Information and Critical Exponents of the Strongly Coupled Plasma}
\author[1,2]{Hajar Ebrahim}
\author[1]{and Gol-Mohammad Nafisi}
\affiliation[1]{Department of Physics, University of Tehran, North Karegar Ave., Tehran 14395-547, Iran}
\affiliation[2]{School of Physics, Institute for Research in Fundamental Sciences (IPM),
	P.O. Box 19395-5531, Tehran, Iran}
\emailAdd{hebrahim@ut.ac.ir}
\emailAdd{gmnafisi@ut.ac.ir}
\abstract{This note contains discussions on the entanglement entropy and mutual information of a strongly coupled field theory with a critical point which has a holographic dual. We investigate analytically, in the specific regimes of parameters, how these non-local operators behave near the critical point. Interestingly, we observe that although the mutual information is constant at the critical point, its slope shows a power-law divergence in the vicinity of the critical point. We show that the leading behavior of mutual information at and near the critical point could yield a set of critical exponents if we regard it as an order parameter. Our result for this set of static critical exponents is $(1/2,1/2,1/2,2)$ which is identical to the one calculated via the thermodynamic quantities. Hence it suggests that beside the numerous merits of mutual information, this quantity also captures the critical behavior of the underlying field theory and it could be used as a proper measure to probe the phase structure associated with the strongly coupled systems.}
\begin{document}
 IPM/P-2020/002
	\maketitle
	\flushbottom
\section{Introduction and Results}
Following the recent advances in theoretical physics, one could observe that the quantum information theory and quantum gravity have become the front-runners of current theoretical research programs. Due to the developments in studying black hole physics via holography in recent years, it has become evident that the concept of entanglement not only plays a crucial role in bringing together these two seemingly unrelated areas, resulting in fruitful insights toward understanding the important properties of underlying physical systems, it might also shed lights on our current view of quantum gravity \cite{Rangamani:2016dms}. For a given bipartitioned system in general, entanglement entropy measures the amount of quantum entanglement between its two sub-systems. In the context of quantum field theory, one could also calculate the entanglement entropy between two spacetime regions using the replica trick method \cite{Holzhey:1994we}. Following up the seminal work of Cardy and Calabrese, in which they obtaind the entanglement entropy of a two dimensional conformal field theory, generalizations of their results for the higher dimensional field theories have been an active line of research \cite{Calabrese:2004eu,Casini:2009sr,Nishioka:2018khk}. It was also shown that the entanglement entropy in field theories suffers from short-distance divergence obeying an area-law behavior which makes it a scheme-dependent quantity in the UV limit \cite{Bombelli:1986rw,Srednicki:1993im}.

In the context of AdS/CFT correspondence \cite{Maldacena:1997re,Gubser:1998bc,Witten:1998qj}, quantum entanglement has become one of the main research interests as well. Ryu and Takayanagi (and later Hubeny, Rangamani and Takayanagi) proposed a general recipe for calculating the entanglement entropy of $d$-dimensional large-N CFTs which admit holographic dual \cite{r01, r02}. Their proposal has successfully satisfied the necessary conditions required for the entanglement entropy of field theories and matched with the prior known results obtained for the two-dimensional CFT \cite{Headrick:2007km, r09, Ryu:2006ef}. The remarkable success of this proposal stimulated numerous works which gave us more insights toward better understanding of this topic \cite{Casini:2011kv, Rahimi:2016bbv, Lokhande:2017jik, Myers:2012ed, Asadi:2018ijf, BabaeiVelni:2019pkw, Fareghbal:2019czx}. 

In order to overcome the scheme-dependent measure of entanglement, one can use a specific linear combinations of entanglement entropies called mutual information which is defined by $I(A:B)\equiv\mathcal{S}(A)+\mathcal{S}(B)-\mathcal{S}(A\cup B)\,$, where $\mathcal{S}$ denotes the entanglement entropy of its associated spacetime region. Mutual information is a finite and positive semi-definite quantity which measures the total correlations between the two disjoint regions $A$ and $B$ \cite{r04, r05, Headrick:2010zt}. We will show that in our background the dominant term in mutual information features an area-law behavior in high temperature limit, in contrast to the entanglement entropy which has a volume-law behavior within the same thermal limit. Therefore mutual information would be a more reliable quantity to be used in order to investigate the physical properties of systems described by QFTs.

In this paper we consider $\mathcal{N}=4$ super Yang-Mills theory at finite temperature, $T$, charged under a $U(1)$ subgroup of its $SU(4)_R$ R-symmetry group which includes one chemical potential, $\mu$, and it is dual to the well-known 1RCBH background \cite{Gubser:1998jb,Behrndt:1998jd,Kraus:1998hv,Cvetic:1999ne,Cvetic:1999rb}. More detailed discussions regarding this background can be found in section \ref{sec01}. Due to the fact that underlying theory is conformal, its phase diagram is one-dimensional and it is characterized by the ratio $\mu / T$. This one-dimensional line ends in a critical point denoted by $\mu_c/T_c=\pi/\sqrt{2}$ \cite{DeWolfe:2010he, r00a}. Since the phase structure of this theory is simple it provides us with an analytically solvable model in order to discuss its critical phenomena in terms of information-theoretic measures such as entanglement entropy and mutual information. We obtain these measures analytically, in the context of gauge/gravity duality, within the various thermal limits and we use mutual information, which is a scheme-independent quantity, as an order parameter and discuss its behavior near the critical point.

Finally by using our results for the holographic mutual information, we obtain the following values for the two suitable independent static critical exponents
\begin{equation}
\delta=2 \qquad \text{and}\qquad \gamma=\frac{1}{2}\;,
\end{equation}
and by using the well-known scaling relations we determine the four static critical exponents to be
\begin{equation}
\left(\alpha,\beta,\gamma,\delta\right)=\left(\frac{1}{2},\frac{1}{2},\frac{1}{2},2\right)\;,
\end{equation}
which are in full agreement with the ones obtained previously in the literature using thermodynamic quantities \cite{r00a, Cai:1998ji,Cvetic:1999rb}.
\section{The Background Geometry}\label{sec01}
As we mentioned in the introduction, we are interested in studying the critical phenomena of a strongly coupled plasma using the framework of holography. Therefore we start with a holographic geometry in five dimensions dual to the aforementioned 4-dimensional field theory with critical point, which is known as the 1RCBH background \cite{Gubser:1998jb,Behrndt:1998jd,Kraus:1998hv,Cvetic:1999ne,Cvetic:1999rb}.

	\subsection{Geometry}
	We consider a gravitational theory on a five dimensional manifold with metric $g_{\mu \nu}$, consisting of a gauge field, $A_\mu\,$, and a scalar field (dilaton), $\phi\,$, which is described by the following Einstein-Maxwell-Dilaton action
	\begin{equation}\label{eq00a}
	\mathcal{S}_{\text{\tiny EMD}}=\frac{1}{16\pi G_N^{(5)}}\int d^5 x\,\sqrt{-g}\left[\mathcal{R}-\frac{f(\phi)}{4} F_{\mu \nu}F^{\mu \nu}-\frac{1}{2}\,(\partial_\mu \phi)(\partial^\mu \phi)-V(\phi) \right]\;,
	\end{equation}
	where $G_N^{(5)}$ is the 5-dimensional Newton constant. The coupling function between the gauge field and the dilaton , $f(\phi)$, and the dilaton potential, $V(\phi)$, are given by\\
	\begin{equation}
	\begin{aligned}
	f(\phi)&=e^{-\sqrt{\frac{4}{3}}\,\phi}\,,\\
	V(\phi)&=-\frac{1}{R^2}\left(8\,e^{\frac{\phi}{\sqrt{6}}}+4\,e^{-\sqrt{\frac{2}{3}}\,\phi} \right)\,,
	\end{aligned}
	\end{equation}
	where $R$ is the asymptotic AdS$_5$ radius. The 1RCBH background is the solution to the equations of motion of the EDM action in eq. (\ref{eq00a}) and it is described by 
	\begin{equation}\label{eq03b}
	ds^2_{(5)}={{e}^{2A(z)}}\left(-h(z)\,dt^2+d{\vec{x}}^{2}_{(3)}\right)+\frac{{{e}^{2B(z)}}}{h(z)}\frac{R^4}{z^4}\,dz^2\;,
	\end{equation}
	where
	\begin{equation}\label{eq03}
	\begin{aligned}
	A(z)&=\ln \left(\frac{R}{z}\right)+\frac{1}{6}\ln \left(1+\frac{Q^2z^2}{R^4}\right)\,, \\
	B(z)&=-\ln \left(\frac{R}{z}\right)-\frac{1}{3}\ln \left(1+\frac{Q^2z^2}{R^4}\right)\,, \\
	h(z)&=1-\frac{M^2\,z^4}{R^6\left(1+\frac{Q^2z^2}{R^4} \right)}\,, \\
	\phi(z)&=-\sqrt{\frac{2}{3}} \ln\left(1+\frac{Q^2z^2}{R^4} \right)\,,\\
	\Phi(z)&=\frac{MQz_h^2}{R^4\left(1+\frac{Q^2z_h^2}{R^4}\right)}-\frac{MQz^2}{R^4\left(1+\frac{Q^2z^2}{R^4}\right)}\;,
	\end{aligned}
	\end{equation}
	in which $\Phi(z)$ is the electric potential given by the temporal component of the gauge field and it is chosen such that it is zero on the horizon and regular on the boundary \cite{DeWolfe:2010he,r00a}. Note that we are working in the Poincare patch coordinates by defining $z=R^2/r$ such that $z$ is the radial bulk coordinate and the boundary lies at $z\rightarrow 0$. The black hole mass is denoted by $M$ while $Q$ denotes its charge. By using the fact that $h(z_h)=0\,$, one can obtain a relation for the mass 
	which then gives us the following expression for the blackening factor
	\begin{equation}\label{eq04} 
	h(z)=1-\left(\frac{z}{z_h}\right)^4\left(\frac{1+\left(\frac{Qz_h}{R^2}\right)^2}{1+\left(\frac{Qz}{R^2}\right)^2} \right)\;.
	\end{equation}
	The location of the black brane horizon, $z_h$, could be expressed in terms of $M$ and $Q$ as
	\begin{equation}\label{eq05}
	z_h=R\,\sqrt{\frac{Q^2+\sqrt{Q^4+4M^2R^2}}{2M^2}}\;.
	\end{equation}
	\subsection{Thermodynamics}\label{sec02}
	The field theory dual to the geometry background discussed in the last subsection is characterized by the temperature, $T\,$, and the chemical potential, $\mu\,$. Following the usual recipe for obtaining the temperature---Wick rotating the temporal coordinate of the metric, performing a Taylor expansion of the metric coefficients near the horizon and imposing the periodicity condition---we obtain the Hawking temperature as
	\begin{equation}\label{eq07a}
	T=\frac{1}{4\pi R^2}\left\lvert e^{A(z_h)-B(z_h)}\,h'(z_h)\,z_h^2\right\rvert \;,
	\end{equation} 
	hence
	\begin{equation}\label{eq07}
	T=\frac{1}{2\pi z_h}\left(\frac{2+\left(\frac{Qz_h}{R^2}\right)^2}{\sqrt{1+\left(\frac{Qz_h}{R^2}\right)^2}}\right)\;,
	\end{equation}
	where the prime symbol in eq. (\ref{eq07a}) denotes the derivative with respect to $z$ coordinate. The chemical potential is given by
	\begin{equation}
	\mu=\frac{1}{R}\lim_{z\to 0}\Phi(z)\;,
	\end{equation}
	therefore
	\begin{equation}\label{eq08}
	\mu=\frac{Q}{R^2\sqrt{1+\left(\frac{Qz_h}{R^2}\right)^2}}\;.
	\end{equation}
	By using eqs. (\ref{eq07}) and (\ref{eq08}) we obtain the following useful non-negative dimensionless quantity
	\begin{equation}\label{eq09}
	\frac{Qz_h}{R^2}=\frac{\sqrt{2}}{\lambda}\left(1\pm\sqrt{1-\lambda^2}\right) \quad s.t.\quad \lambda\equiv \left(\frac{\mu/T}{\pi/\sqrt{2}}\right)\;.
	\end{equation}
	For our future use, we rewrite temperature in terms of the dimensionless quantity $Qz_h/R^2$ as  
\begin{equation}\label{temp}
T={\hat{T}}\left(\frac{1+\frac{\xi}{2}}{\sqrt{1+\xi}}\right)\;,
	\end{equation} 
where we have defined ${\hat{T}}\equiv1/{\pi z_h}$ and $\xi\equiv Q^2z_h^2/R^4\,$. In order to see which sign of the eq. (\ref{eq09}) relates to a thermodynamically stable phase, one needs to obtain the entropy and charge density in terms of $\mu/T$ first. One can show that the entropy density, $s$, and $U(1)$ charge density, $\rho$, are given by
	\begin{equation}
	s=\frac{R^3}{4G_N^{(5)}z_h^3}\,\sqrt{1+\xi} \qquad,\qquad \rho=\frac{QR}{8\pi G_N^{(5)} z_h^2}\,\sqrt{1+\xi}\;.
	\end{equation}
	Now suppose that the thermodynamic potential of a system is given by $\Phi(x_1,...,x_r)\,$ depending on some set of variables $\{x_1,...,x_r\}\,$. Then for a stable phase, the Hessian matrix, \textbf{H}, of the associated potential defined by
	\begin{equation}
	\mathbf{H}_{ij}\equiv\left[\frac{\partial^2\Phi}{\partial x_i\,\partial x_j} \right]\;,
	\end{equation}
	should be positive-definite\footnote{Note that the converse does not necessarily imply the global stability since the positive-definiteness of Hessian matrix for a convex function indicates a local minima, therefore the stability should be considered a local one instead.}.
	Here we can choose the free energy density $f$ which satisfies $\,-df=sdT+\rho d\mu\,$, as our relevant thermodynamic potential. Hence by evaluating its Hessian matrix which then reduces to  $\,\mathbf{H}= \left[\partial (s,\rho)/\partial (T,\mu)\right]\,$, we find out that if we choose the minus sign in eq. (\ref{eq09}), both principal minors of $\mathbf{H}$ become strictly positive for $\mu/T\in [0,\pi/\sqrt{2}]\,$ or $\lambda \in [0,1]\,$. Therefore \textbf{H} is positive-definite and the local thermodynamic stability of the field theory dual to 1RCBH background is guaranteed. Note that since $\lambda \in [0,1]\,$ then the parameter $Qz_h/R^2\in [0,\sqrt{2}]\,$, therefore $\xi$, would be a number of the one order of magnitude.
	
	In order to classify the phase transitions in this model, we observe that for the second derivatives of the free energy density with respect to $T$ and $\mu$ we have
	\begin{equation}\label{sus}
	-\left(\frac{\partial^2 f}{\partial T^2} \right)_{\mu}=\left(\frac{\partial s}{\partial T} \right)_\mu\equiv \frac{C_\mu}{T}\qquad \text{and} \qquad -\left(\frac{\partial^2 f}{\partial \mu^2} \right)_T=\left(\frac{\partial \rho}{\partial \mu} \right)_T\equiv \chi_2\;,
	\end{equation}
	where $C_\mu$ is the specific heat at constant chemical potential and $\chi_2$ is the 2nd order R-charge susceptibility. One could see that both diverge at $\mu/T=\pi/\sqrt{2}$ and thermodynamic quantities of the 1RCBH background will end at the point $\mu_c/T_c=\pi/\sqrt{2}\,$ or equivalently at $Qz_h/R^2=\sqrt{2}\,$. In other words, the phase structure of the field theory dual to this background will exhibit a second order phase transition and the critical point is characterized by the ratio $\mu/T$ as expected, since the underlying theory is conformal.
	 
	\section{Holographic Entanglement Entropy}\label{sec03}
	Suppose that a CFT exists on a Cauchy surface $\mathcal{C}$ of a $d$-dimensional Lorentzian manifold $\mathcal{B}_d\,$. We define region $A$ to be a subset of $\mathcal{C}$ such that $A\cup A^c=\mathcal{C}$ where $A^c$ is its complement. This region has a boundary $\partial A\,$ (the entangling surface) which is a co-dimension 2 hypersurface in $\mathcal{B}_d\,$. We then assume that the Hilbert space $\mathcal{H}$ of the CFT can be factorized into $\mathcal{H}_A\otimes \mathcal{H}_{A^c}$ and we let $\rho$ to be a density operator (matrix) associated to a state $\ket{\psi}\in \mathcal{H}\,$. Now by defining the reduced density operator for region $A$ to be $\rho_A\equiv \text{tr}_{A^c}(\rho)$ where $\text{tr}_{A^c}$ denotes the partial trace over $A^c\,$, one can measure the entanglement between regions $A$ and $A^c$ using the von Neumann entropy\footnote{By assuming that this measure is mathematically well-defined in QFT.} which is a non-local quantity defined by
	\begin{equation}
	\mathcal{S}(A)\equiv -\text{tr}(\rho_A\,\log \rho_A)\;.
	\end{equation}
	In the framework of the AdS/CFT correspondence where we have a $d$-dimensional CFT dual to a $(d+1)$-dimensional asymptotically AdS spacetime $\mathcal{M}_{d+1}$, one can use the holographic entanglement entropy (Ryu-Takayanagi and Hubeny-Rangamani-Takayanagi prescriptions) which is given by \cite{r01, r02}
	\begin{equation}\label{eq00}
	\mathcal{S}(A)=\frac{\mathcal{A}(\gamma_A)}{4\,G_N^{(d+1)}}\;,
	\end{equation} 
	where $\gamma_A$ is a co-dimension 2 extremal surface in $\mathcal{M}_{d+1}$ with the area $\mathcal{A}(\gamma_A)$ such that $\partial \gamma_A=\partial A$ and $G_N^{(d+1)}$ is the $(d+1)$-dimensional Newton constant. This recipe has already passed the tests one expects for the entanglement entropy. Also the quantities derived from this relation such as holographic mutual information satisfy all the necessary conditions---as well as an extra feature called monogamy--- required for any entanglement measure in the context of quantum information theory \cite{Headrick:2007km, r09, r08, Headrick:2010zt}. 
	
	\subsection{Set up}
\begin{figure}[t]
		\centering
		\includegraphics[width=0.7\textwidth]{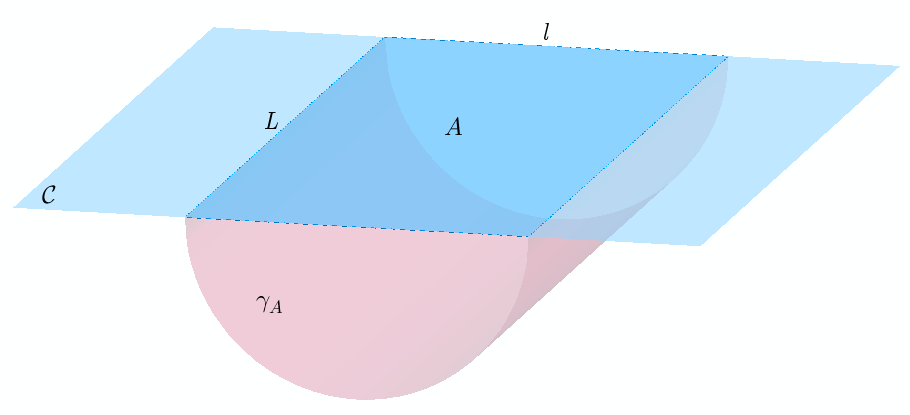}
		\caption{A simplified sketch of a strip $A$ on the Cauchy surface $\mathcal{C}$ with characteristic length $l$ which has a unique minimal surface $\gamma_A$ in the bulk anchored on its boundary. }
		\label{fig01}
	\end{figure}

	In the holographic set up, we choose our boundary system to be an infinite rectangular strip of characteristic length $l$ (Fig.\ref{fig01}) and we parameterize the boundary coordinate $x$ in terms of the bulk coordinate $z$. We specify this strip by
	\begin{equation}
	x^{(1)}\equiv x \in [-\frac{l}{2},\frac{l}{2}] \quad,\quad x^{(i)}\in [-\frac{L}{2},\frac{L}{2}]\; ,\; i=2,3\;,
	\end{equation}
	such that $L\to \infty\,$.
	\noindent
	\subsection{Area and Characteristic Length}\label{subsec01}
	\noindent
	For a general bulk manifold $\mathcal{M}_{d+1}$ with the metric $g_{\mu\nu}\,$, the extremal surface $\gamma_A$ is a co-dimension 2 hypersurface in $\mathcal{M}_{d+1}$ whose area functional is given by
	\begin{equation}\label{eq10}
	\mathcal{A}(\gamma_A)=\int d^{d-1}x\,\sqrt{\det \left(g_{ M N}\right)}\;,
	\end{equation} 
	where $g_{MN}$ is the induced metric on $\gamma_A$. For the geometric background of eq. (\ref{eq03}) on the constant time slice, we parameterize $x\equiv x(z)$ and obtain the area as
	\begin{equation}\label{eq11}
	\mathcal{A}=2L^2\int_{0}^{z_c} dz\,e^{3A(z)}\,\sqrt{x'(z)^2+\frac{R^4}{z^4\,h(z)}\,e^{2\left(B(z)-A(z)\right)}}\;.
	\end{equation}
	Since the integrand of eq. (\ref{eq11}) does not have an explicit dependence on $x$, if we construct its Hamiltonian we get the following differential equation
	\begin{equation}\label{eq12}
	x'(z)\equiv\frac{dx}{dz}=\frac{R^2}{z^2}\frac{e^{3A(z_c)}\,e^{B(z)-A(z)}}{\sqrt{h(z)}\sqrt{e^{6A(z)}-e^{6A(z_c)}}}\;,
	\end{equation}
	where $z=z_c$ is the extrema of the minimal surface where $z'(x)=0$. By substituting eq. (\ref{eq12}) in eq. (\ref{eq11}) we obtain
	\begin{equation}\label{eq13}
	\begin{aligned}
	\mathcal{A}&=2L^2R^3\int_{0}^{z_c}dz\,\frac{z_c^3}{z^6}\sqrt{\frac{1+\xi\left(\frac{z}{z_h}\right)^2}{1+\xi\left(\frac{z_c}{z_h}\right)^2}}\left[1-\left(\frac{z}{z_h}\right)^4\left(\frac{1+\xi}{1+\xi\left(\frac{z}{z_h}\right)^2} \right)\right]^{-\frac{1}{2}}\\
	&\qquad\qquad\qquad\quad\quad\,\times\left[\left(\frac{z_c}{z}\right)^6\left(\frac{1+\xi\left(\frac{z}{z_h}\right)^2}{1+\xi\left(\frac{z_c}{z_h}\right)^2}\right)-1\right]^{-\frac{1}{2}}\;,
	\end{aligned}
	\end{equation}
	where we have used the definition $\xi\equiv\left(Qz_h/R^2\right)^2\,$ which we introduced previously in section \ref{sec02}. By integrating the differential equation of eq. (\ref{eq12}) and imposing the boundary conditions $x(z_c)=0$ and $x(0)=\pm\, l/2\,$, we obtain the following expression for the characteristic length
	\begin{equation}\label{eq14}
	\begin{aligned}
	\frac{l}{2}&=\int_{0}^{z_c}dz\,\left[1+\xi\left(\frac{z}{z_h}\right)^2\right]^{-\frac{1}{2}}\left[1-\left(\frac{z}{z_h}\right)^4\left(\frac{1+\xi}{1+\xi\left(\frac{z}{z_h}\right)^2} \right)\right]^{-\frac{1}{2}}\\
	&\qquad\qquad\,\,\times\left[\left(\frac{z_c}{z}\right)^6\left(\frac{1+\xi\left(\frac{z}{z_h}\right)^2}{1+\xi\left(\frac{z_c}{z_h}\right)^2}\right)-1\right]^{-\frac{1}{2}}\;.
	\end{aligned}
	\end{equation}
	Since it is not easy to calculate this integral analytically, by the help of the generalized multinomial expansions given in the appendix \ref{app1} we show that the eq. (\ref{eq14}) can be represented by the following series\footnote{This method of calculating the entanglement entropy and mutual information has been initially used in \cite{Fischler:2012ca, r06, r07}.} 
	\begin{equation}\label{eq15}
	\frac{l}{2}=z_c\,\sum_{k=0}^{\infty}\sum_{n=0}^{k}\sum_{m=0}^{\infty}\sum_{j=0}^{\infty}\, G_{knmj}\,F_{knmj} \left(\frac{z_c}{z_h}\right)^{2(k+n+m)}\;,
	\end{equation}
	where
	\begin{equation}\label{eq16}
	\addtolength{\jot}{1em}
	\begin{aligned}
	G_{knmj}&\equiv \frac{\Gamma\left(k+\frac{1}{2}\right)\,\Gamma\left(j+m+\frac{1}{2}\right)\,\Gamma\left(2+3j+k+n\right)}{2\pi\,\Gamma\left(n+1\right)\,\Gamma\left(k-n+1\right)\,\Gamma\left(j+1\right)\,\Gamma\left(3+3j+k+n+m\right)}\,\,,\\
	F_{knmj}&\equiv (-1)^{k+n}\, \xi^{k-n+m}\,\left(1+\xi\right)^n\left[1+\xi\left(\frac{z_c}{z_h}\right)^2\right]^{-m}\;.
	\end{aligned}
	\end{equation}
	Note that in order to make use of the binomial expansions for negative powers we made sure that the following relations are satisfied for the whole range of $\xi\in [0,2]$ and for $z_c$ between boundary and the horizon
	\begin{equation}
	\frac{\xi\left(\frac{z_c}{z_h}\right)^2}{1+\xi\left(\frac{z_c}{z_h}\right)^2}\left(1-\frac{z^2}{{z_c}^2}\right) < 1\qquad {\text{and}}\qquad \xi\left(\frac{z}{z_h}\right)^2-(1+\xi)\left(\frac{z}{z_h}\right)^4 <1\;.
	\end{equation}
	These expansions can also be used to represent the area in eq. (\ref{eq13}) by
	\begin{equation}\label{eq16a}
	\begin{aligned}
	\mathcal{A}&=\frac{2L^2R^3}{\pi}\sum_{k=0}^{\infty}\sum_{n=0}^{k}\sum_{m=0}^{\infty}\sum_{j=0}^{\infty} \frac{\Gamma\left(k+\frac{1}{2}\right)\,\Gamma\left(j+m+\frac{1}{2}\right)}{\Gamma\left(n+1\right)\,\Gamma\left(k-n+1\right)\,\,\Gamma\left(j+1\right)\,\Gamma\left(m+1\right)\,}\\
	&\quad\quad\quad\quad\,\times (-1)^{k+n}\,\xi^{k-n+m}\,\left(1+\xi\right)^n\,\left[1+\xi\left(\frac{z_c}{z_h}\right)^2\right]^{-m-\frac{1}{2}}\left(\frac{z_c}{z_h}\right)^{2m}\\
	&\quad\quad\quad\quad\,\times \int_{0}^{z_c} dz\,\left[1+\xi\left(\frac{z}{z_h}\right)^2\right]\left[1-\left(\frac{z}{z_c}\right)^2\right]^m\,z^{-3}\left(\frac{z}{z_c}\right)^{6j}\left(\frac{z}{z_h}\right)^{2(k+n)}\;.
	\end{aligned}
	\end{equation}
	As one would expect in general, the area enclosed by the extremal surface is divergent due to its near boundary behavior. Here one could show that the last integral (hence the area) remains finite if the condition $k+n+3j>1$ is satisfied. Hence we need to isolate $(k=n=j=0)$ and $(k=1,\,n=j=0)$ terms together and perform their sum over $m$ to get the part of the area in which the divergent term is contained. By doing so, we obtain
	\begin{equation}\label{eq16b}
	\mathcal{A}_0\equiv L^2R^3\left\{\frac{1}{\epsilon^2}+\frac{3\,\xi}{2z_h^2}-\frac{1}{z_c^2}\left[1+\xi\left(\frac{z_c}{z_h}\right)^2\right]^{\frac{3}{2}} \right\}\;,
	\end{equation}
	where $z=\epsilon$, such that $\epsilon\to 0$, is the cut-off surface in the bulk geometry related to the UV regulator of the field theory. We see that the divergent term in eq. (\ref{eq16b}) has an area-law behavior which appears in the corresponding holographic entanglement entropy as well. This result is indeed expected in a $d$-dimensional field theory side where the leading divergence in the UV limit $\epsilon\to 0$ obeys an area-law. For convenience, we will work with the finite part of the area henceforth by subtracting the $1/\epsilon^2$ term\footnote{Note that our preferred cut-off independent measure of entanglement would be the mutual information instead, as we discuss it in section \ref{sec04}.}. It is given by
	\begin{equation}\label{eq21}
	\addtolength{\jot}{1em}
	\begin{aligned}
	\mathcal{A}_{\text{\tiny fin}}&=\frac{L^2R^3}{z_c^2}\left\{\frac{3\,\xi}{2}\left(\frac{z_c}{z_h}\right)^2-\left[1+\xi\left(\frac{z_c}{z_h}\right)^2\right]^{\frac{3}{2}}+\frac{1+\xi}{3\,\xi }\left(\frac{z_c}{z_h}\right)^2\left[\left(1+\xi\left(\frac{z_c}{z_h}\right)^2\right)^{\frac{3}{2}}-1\right]\right\}\\
	&+\frac{L^2R^3}{z_c^2}\Biggl\{\sum_{k=2}^{\infty}\sum_{n=0}^{k}\sum_{m=0}^{\infty}\,\Lambda_{knm}\,\frac{\Gamma\left(m+\frac{1}{2}\right)\,\Gamma\left(k+n-1\right)\,}{\Gamma\left(k+n+m+1\right)\,}\left(\frac{z_c}{z_h}\right)^{2(k+n+m)}\\
	&\qquad\qquad\qquad\qquad\quad\,\,\times\left[\left(m+1\right)+\left(k+n-1\right)\left(1+\xi\left(\frac{z_c}{z_h}\right)^2\right)\right]\Biggr\}\\
	&+\frac{L^2R^3}{z_c^2}\Biggl\{\sum_{k=0}^{\infty}\sum_{n=0}^{k}\sum_{m=0}^{\infty}\sum_{j=1}^{\infty}\,\Lambda_{knm}\,\frac{\Gamma\left(m+j+\frac{1}{2}\right)\,\Gamma\left(k+n+3j-1\right)\,}{\Gamma\left(j+1\right)\,\Gamma\left(k+n+m+3j+1\right)\,}\left(\frac{z_c}{z_h}\right)^{2(k+n+m)}\\
	&\qquad\qquad\qquad\qquad\qquad\,\,\,\,\,\times\left[\left(m+1\right)+\left(k+n+3j-1\right)\left(1+\xi\left(\frac{z_c}{z_h}\right)^2\right)\right]\Biggr\}\;,
	\end{aligned}
	\end{equation}
	where
	\begin{equation}
	\Lambda_{knm}\equiv \frac{(-1)^{k+n}\,\Gamma\left(k+\frac{1}{2}\right)}{\pi\,\Gamma\left(n+1\right)\,\Gamma\left(k-n+1\right)}\,\xi^{k-n+m}\,\left(1+\xi\right)^n\,\left[1+\xi\left(\frac{z_c}{z_h}\right)^2\right]^{-m-\frac{1}{2}}\;.
	\end{equation}
	We should point out that although this result for the area is lengthy and hard to work with, it gives us the vantage point of investigating the behaviors of entanglement entropy and mutual information near the critical point analytically, which we will discuss in the forthcoming sections. 
 
	\section{Entanglement Entropy and Thermal Limits}\label{sec001}
	As one could observe in eq. (\ref{eq21}), the area of the minimal surface would be characterized by its two dimensionless parameters $\xi$ and $z_c/z_h\,$. In this section we investigate the holographic entanglement entropy with respect to $z_c/z_h$ which introduces two thermal limits, while we leave its analysis with regard to the parameter $\xi\,$ which controls the critical behavior to section \ref{sec05}. Now given the ratio of the extremal surface location to the horizon location, i.e. $z_c/z_h$, one could expect to see two different cases for the area obtained in the previous section (hence for the entanglement entropy) namely when $z_c/z_h\ll 1$ and when we have $z_c/z_h\sim 1\,$. Note that the former implies that the minimal surface is near the AdS boundary while latter indicates the case where minimal surface approaches the horizon while never penetrating it. This is due to the fact that in a static asymptotically AdS spacetime, the minimal surface does not pass beyond the horizon of an existing black hole \cite{r13}\footnote{We will comment on this point in appendix \ref{app4} where we show how close could minimal surface get to the horizon in the high temperature limit.}. For the field theory side with the introduced scale $l\,$, we can immediately translate the aforementioned cases into the two inequivalent thermal limits; ${\hat{T}}l\ll 1$ and ${\hat{T}}l\gg 1$, respectively where ${\hat{T}}$ is defined in eq. (\ref{temp}). Hence one could identify the $z_c/z_h\ll 1$ case with the low temperature limit associated with the ground state fluctuations of CFT while the $z_c/z_h\sim 1$ case could be identified with the high temperature limit in which the entanglement of the thermal excitations is considered.  
	
	\subsection{Low Temperature Case}
	One of the main concerns while dealing with the infinite series representation of functions is the issue of their convergence, since depending on their growth, they might simply diverge as well. In the low temperature limit where $z_c/ z_h\ll 1\,$, we observe that both infinite series in eqs. (\ref{eq15}) and (\ref{eq21}) converge. Therefore we can expand eq. (\ref{eq15}) at 4th order in $(z_c/z_h)$ obtaining
	\begin{equation}\label{eq17}
	l=z_c\left\{a_1-\frac{a_1\,\xi}{6}\left(\frac{z_c}{z_h}\right)^2+\left[\frac{a_2\,(1+\xi)}{2}+\frac{a_3\,\xi^2}{24} \right]\left(\frac{z_c}{z_h}\right)^4+\mathcal{O}\left(\frac{z_c}{z_h}\right)^6 \right\}\;,
	\end{equation}
	where we performed the sum over $j$ and the numerical constants $a_1,\,a_2$ and $a_3$ are given in the appendix \ref{app2}. By solving eq. (\ref{eq17}) perturbatively for $z_c$ at 4th order in $\left(l/z_h\right)$ we get
	\begin{equation}\label{eq20}
	z_c=\frac{l}{a_1}\left\{1+\frac{\xi}{6a_1^2}\left(\frac{l}{z_h}\right)^2+\frac{1}{2a_1^4}\left[\frac{\xi^2}{6}\left(1-\frac{a_3}{2 a_2}\right)-\frac{a_2}{a_1}\left(1+\xi\right)\right]\left(\frac{l}{z_h}\right)^4+\mathcal{O}\left(\frac{l}{z_h}\right)^6  \right\}\;.
	\end{equation}
	Now if we expand the finite part of the area in eq. (\ref{eq21}) to the lowest orders, we obtain
	\begin{equation}\label{eq23}
	\addtolength{\jot}{1em}
	\begin{aligned}
	\mathcal{A}^{\text{\tiny finite}}_{\text{\tiny low}}&=\frac{L^2R^3}{z_c^2}\left[\frac{1+\xi}{2}\left(\frac{z_c}{z_h}\right)^4-1\right]\\
	&+\frac{L^2R^3}{z_c^2}\sum_{j=1}^{\infty}\frac{\Gamma\left(j+\frac{1}{2}\right)}{\sqrt{\pi}\,\Gamma\left(j+1\right)\,(3j-1)}\Biggl[1+\frac{\xi}{3}\left(\frac{z_c}{z_h}\right)^2\\
	&\quad\quad\quad\quad\quad\quad\quad\quad\quad\quad\quad\quad\quad\quad\quad+\left(\frac{\left(-4\, \xi ^2+9\, \xi +9\right) j-3 (\xi +1)}{18 j+6}\right)\left(\frac{z_c}{z_h}\right)^4 \Biggr]\;.
	\end{aligned}
	\end{equation}
	Finally, by performing the sum and substituting for $z_c$ from eq. (\ref{eq20}) in the last expression and then using eq. (\ref{eq00}), we obtain the entanglement entropy in the low temperature limit as
	\begin{equation}\label{eq24}
	\begin{aligned}
	\mathcal{S}^{\text{\tiny finite}}_{\text{\tiny low}}=\frac{R^3}{4G_N^{(5)}}\left(\frac{L}{l}\right)^2&\Bigg\{a_1^2\,(w_1-1)+\frac{\xi}{3}\left(\frac{l}{z_h}\right)^2\\
	&+\frac{1}{2 a_1^2}\Bigg[(1+\xi)\big(1-w_3+3 w_2+\frac{2\,(w_1-1)\,a_2}{a_1}\big)\\
	&\qquad\quad+\frac{\xi^2}{6}\big((w_1-1)(\frac{a_3}{a_1}-1)-8w_2\big)\Bigg]
	\left(\frac{l}{z_h}\right)^4 \Bigg\},
	\end{aligned}
	\end{equation}
	where the numerical constants $w_1,\,w_2$ and $w_3$ are given in the appendix \ref{app2}. We note that in the limit where $Q\rightarrow 0\,$, we get $z_h=1/\pi T$ and the subleading terms become 2nd and 4th order in $Tl$ as expected from the AdS-RN results. To make this relation more transparent we define 
	\begin{equation}\label{eqc}
	\begin{aligned}
	c&\equiv a_1^2\,(w_1-1) \approx -0.32 \;,\\
	f(\xi)&\equiv(1+\xi)\, \frac{\big(1-w_3+3 w_2+2\,(w_1-1)(\frac{\,a_2}{a_1})\big)}{a_1^2}+\frac{\xi^2}{6}\,\frac{\big((w_1-1)(\frac{a_3}{a_1}-1)-8w_2\big)}{a_1^2}\\ &\approx 1.13\, (1+\xi)- 1.43\, (\frac{\xi^2}{6})\;.\\
	\end{aligned}
	\end{equation}	
	
	The first term in eq. (\ref{eq24}) which we denoted by $c$ in the last expression, does not depend on temperature and it is the contribution of the AdS boundary. Another consistency check for our result would be the case in which we set the chemical potential to zero. The metric of the 1RCBH background then reduces to the AdS-Schwarzschild metric and it is easy to see that we recover the result which was obtained previously in the literature for this particular background \cite{Fischler:2012ca}. 
	
	By using the reparametrization of eq. (\ref{temp}) we can rewrite the low temperature limit of entanglement entropy as
	\begin{equation}\label{low}
	\mathcal{S}^{\text{\tiny finite}}_{\text{\tiny low}}=\frac{R^3}{4G_N^{(5)}}\left(\frac{L}{l}\right)^2 \Bigg\{c +\frac{\xi}{3}\,(\pi {\hat{T}}l)^2
	 +\frac{1}{2}\, f(\xi)\, (\pi {\hat{T}}l)^4 \Bigg\}\;,
	\end{equation}
where ${\hat{T}}$ would be equal to $T$ in the limit $Q\rightarrow 0\,$. The dependence on $\xi$,  which would appear in the mutual information as well, will be utilized later in order to investigate its behavior near the critical point. 

	\subsection{High Temperature Case}
	As we mentioned in the previous section, infinite series do not always converge. Fortunately, for a given divergent series some methods of summability or regularization are available to apply in order to overcome the issue of divergence. We observe that in the high temperature limit where $z_c\sim z_h\,$, the infinite sum of eq. (\ref{eq21}) does not converge\footnote{This divergence is due to the growth of series for $z_c=z_h\,$ and it is not related to UV divergence.}. By making use of the mentioned methods however, we can regularize this series and make it convergent by rearranging it in such a way that we could recover a term proportional to $l\,$\footnote{We will show in appendix \ref{app4} that the sum for $l$ in eq. (\ref{eq15}) converges for $z_c\sim z_h\,$ after regularization.}. We have included the full expression of the resulted regularized series in the appendix \ref{app3}. Therefore we can take the limit $z_c\to z_h$ of eq. (\ref{eq19b}) and by using eq. (\ref{eq00}), we obtain the entanglement entropy in the high temperature regime as
	\begin{equation}\label{eq25}
	\mathcal{S}^{\text{\tiny finite}}_{\text{\tiny high}}=\frac{R^3}{4G_N^{(5)}}\left(\frac{L}{z_h}\right)^2\Bigg\{\sqrt{1+\xi}\left(\frac{l}{z_h}\right)+ \left(\mathcal{S}_1+\mathcal{S}_2+\mathcal{S}_3\right)\Bigg\}\;,
	\end{equation}
	where we defined
	\begin{equation}\label{eq26b}
	\addtolength{\jot}{1em}
	\begin{aligned}
	\mathcal{S}_1&\equiv \frac{3\,\xi}{2}-\frac{1}{3}-\frac{11}{5\, \xi }-\frac{244}{105\, \xi ^2}-\frac{32}{35\, \xi ^3}-\frac{16}{35\, \xi ^4}\\
	&+\sqrt{\xi +1} \left(-\frac{64\,\xi}{105}-\frac{124}{105}+\frac{26}{21\, \xi }+\frac{214}{105\, \xi ^2}+\frac{24}{35\, \xi^3}+\frac{16}{35\, \xi ^4}\right)\,,\\
	\mathcal{S}_2&\equiv \sum_{k=2}^{\infty}\sum_{n=0}^{k}\sum_{m=0}^{\infty}\frac{\Gamma\left(k+\frac{1}{2}\right)\,\Gamma\left(m+\frac{1}{2}\right)\,\Gamma\left(k+n+2\right)\,(-1)^{k+n}\,\xi^{k-n+m}\,(1+\xi)^{n-m-\frac{1}{2}}}{\pi \,\Gamma\left(n+1\right)\,\Gamma\left(k-n+1\right)\,\Gamma\left(k+n+m+3\right)\,}\\
	&\quad\times \left\{\frac{m+1}{k+n-1}\left[1+\frac{m+1}{k+n}\left(2+\frac{m}{k+n+1}\right)\right]+\frac{(1+\xi)(m+1)}{k+n}\left(2+\frac{m}{k+n+1}\right) \right\}\,,\\
	\mathcal{S}_3&\equiv
	\sum_{k=2}^{\infty}\sum_{n=0}^{k}\sum_{m=0}^{\infty}\sum_{j=1}^{\infty}
	\frac{\Gamma\left(k+\frac{1}{2}\right)\,\Gamma\left(j+m+\frac{1}{2}\right)\,\Gamma\left(k+n+3j+2\right)}{\pi \,\Gamma\left(n+1\right)\,\Gamma\left(j+1\right)\,\Gamma\left(k-n+1\right)\,\Gamma\left(k+n+m+3j+3\right)\,}\\
	&\quad\times (-1)^{k+n}\,\xi^{k-n+m}\,(1+\xi)^{n-m-\frac{1}{2}}\\
	&\quad\times \Biggl\{\frac{m+1}{k+n+3j-1}\left[1+\frac{m+1}{k+n+3j}\left(2+\frac{m}{k+n+3j+1}\right)\right]\\
	&\quad\quad\quad+\frac{(1+\xi)(m+1)}{k+n+3j}\left(2+\frac{m}{k+n+3j+1}\right) \Biggr\}\,.
	\end{aligned}
	\end{equation}
	By using eq. (\ref{temp}) we obtain  
		\begin{equation}\label{eq26c}
	\mathcal{S}^{\text{\tiny finite}}_{\text{\tiny high}}=\frac{R^3}{4G_N^{(5)}}\left(\frac{L}{l}\right)^2\left\{ \sqrt{1+\xi}\; (\pi\hat{T}l)^3 + \mathcal{S}_4\,(\pi\hat{T}l)^2 \right\}\;,
	\end{equation}
	where we defined $\mathcal{S}_4\equiv \mathcal{S}_1+\mathcal{S}_2+\mathcal{S}_3\,$, for convenience. We note that the finite leading temperature dependent term (first term in eq. (\ref{eq26c})) scales with the volume of the rectangular strip, $L^2 l$, while the sub-leading term is area dependent. Hence the first term describes the thermal entropy while the second term corresponds to the entanglement entropy between the strip region and its complement, and within this thermal limit the largest contribution comes from the near horizon part of the minimal surface. 
	
	\section{Holographic Mutual Information}\label{sec04}
	We mentioned in the section \ref{subsec01} that the area of an extremal surface has a divergent nature in general and it needs to be regulated. This fact immediately implies the dependency of the holographic entanglement entropy to the choice of a cut-off hypersurface near the boundary. To avoid a regulator-dependent measure of entanglement, one could borrow another quantity from quantum information theory called the mutual information which is a well-defined entanglement  measure in the context of QFT \cite{r04}. For given disjoint regions $A,B\subset \mathcal{C}\,$, the mutual information is defined by\footnote{Simply, $I(A:B)$ quantifies the amount of common information between $A$ and $B\,$.}
	\begin{figure}[t]
		\centering
		\includegraphics[width=0.6\textwidth]{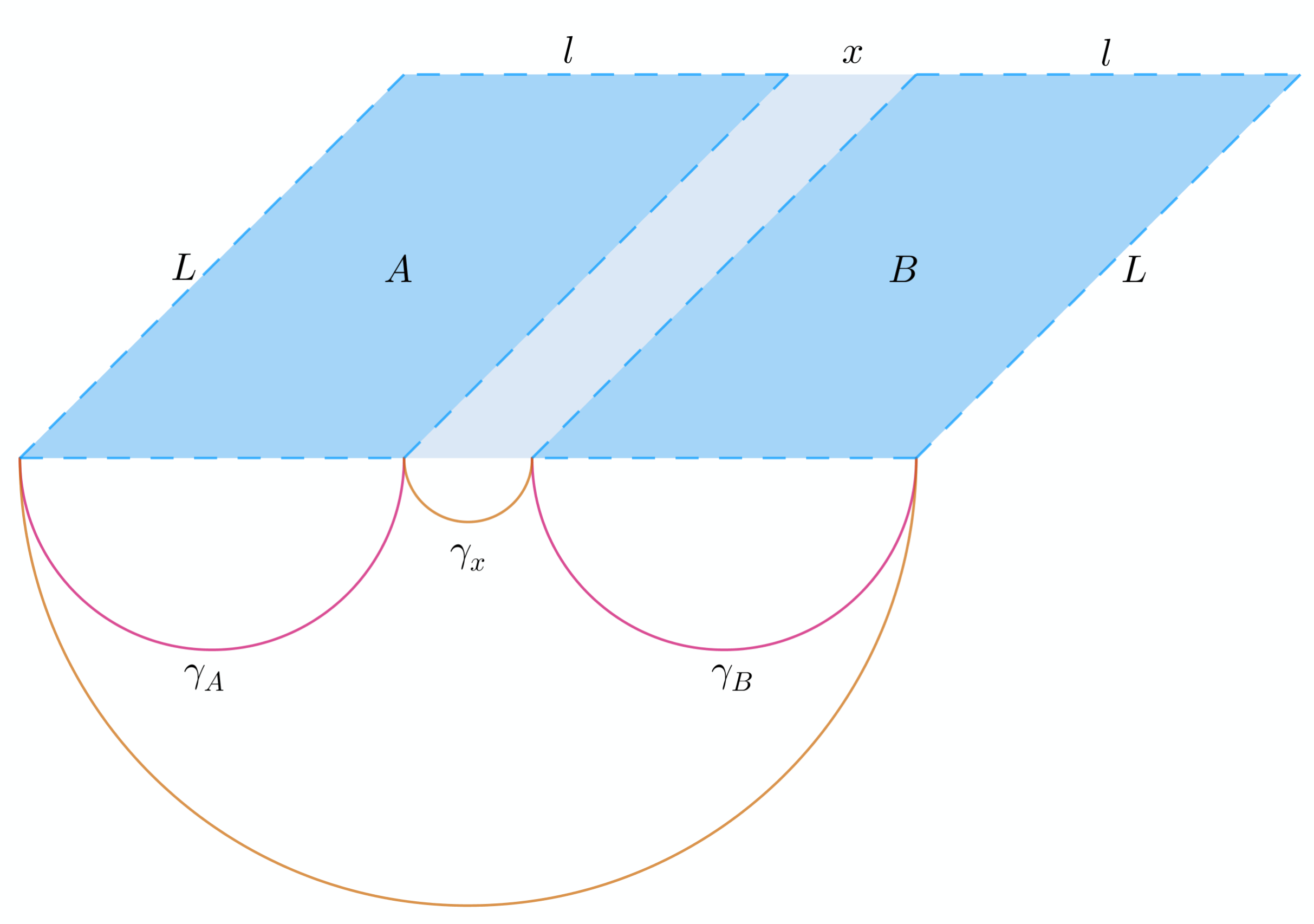}
		\caption{A naive sketch of the case where two disjoint strips $A$ and $B$ are separated by the distance $x$ with the choices for minimal surfaces. The union of brown curves represents the choice of minimal surface for $A\cup B$ when the separation distance is small enough.}
		\label{fig02}
	\end{figure}
	\noindent
	\begin{equation}\label{eq27}
	I(A:B)=\mathcal{S}(A)+\mathcal{S}(B)-\mathcal{S}(A\cup B)\;,
	\end{equation}
	where $\mathcal{S}(A\cup B)$ denotes the entanglement entropy of the composite region $\rho_{AB}\,$. First we note that this measure is positive-semidefinite, since by using the subadditivity inequality of the von Neumann entropy which states that $\mathcal{S}(A)+\mathcal{S}(B)\ge \mathcal{S}(A\cup B)$, one can easily show that $I(A:B)\ge 0$ where the equality is satisfied if the density matrix of the composite region factorizes as $\rho_{AB}=\rho_A\otimes \rho_B\,$. It was also shown that mutual information incorporates the total amount of correlations between two subsystems or equivalently two separate spacetime regions $A$ and $B$ \cite{r05}. More importantly, mutual information is regulator-independent since the UV divergences of $\mathcal{S}(A)$ and $\mathcal{S}(B)$ are canceled by those in $\mathcal{S}(A\cup B)\,$. 
	
	In our set up, we let the two disjoint systems both be infinite rectangular strips of size $l$  which are separated by the distance $x$ on the boundary (Fig.\ref{fig02}). For the minimal surface $\gamma_{A\cup B}\,$, satisfying the condition $\partial \gamma_{A\cup B}=\partial (A\cup B)\,$, we have two choices: when the separation distance is large enough, one can deduce that the $\mathcal{A}(\gamma_{A\cup B})>\mathcal{A}(\gamma_A \cup \gamma_B)$ hence it follows that one would have $\mathcal{S}(A\cup B)=\mathcal{S}(A)+\mathcal{S}(B)$ which then results in the vanishing mutual information \cite{Headrick:2010zt}. On the other hand when $x$ is small enough, $\mathcal{A}(\gamma_{A\cup B})$ would be equal to $\mathcal{A}(\gamma_x)$ plus the area of the minimal surface corresponding to the entire union of the regions $A,\,B$ and $x\,$. Therefore one can assume that there would be a critical separation distance larger than which the mutual information vanishes and the two regions $A$ and $B$ become disentangled. This has been shown in \cite{Headrick:2010zt}. For the non-vanishing mutual information we have
	\begin{equation}\label{eq29}
	I(A:B)=2\,\mathcal{S}(l)-\mathcal{S}(x)-\mathcal{S}(2l+x)\;.
	\end{equation}
	We will use this relation to discuss the behavior of mutual information in different thermal limits. 

	\subsection{Mutual Information and Thermal Limits}\label{mlimits}
	Since the mutual information is a linear combination of entanglement entropies, one could similarly investigate its behavior with respect to the thermal limits which we discussed in  section \ref{sec001}. In addition to those cases, we are able to compare the location of horizon to the newly introduced separation distance as well, which would be specified by the dimensionless ratio $x/z_h\,$. In the field theory, it would mean that the parameter ${\hat{T}}x\,$ introduces an extra temperature limit. Therefore we identify $(l/z_h\ll 1) \wedge (x/z_h\ll 1)$ or $({\hat{T}}l\ll 1 ) \wedge ({\hat{T}}x\ll 1)$ with the low temperature case, whereas $(x/z_h\ll 1) \wedge (l/z_h \gg 1)$ or $({\hat{T}}x\ll 1) \wedge ({\hat{T}}l\gg 1 )$ identifies the additional intermediate temperature limit and finally $(x/z_h\gg 1) \wedge (l/z_h \gg 1)$ or $({\hat{T}}x\gg 1) \wedge ({\hat{T}}l\gg 1 )$ characterizes the high temperature regime where ${\hat{T}}$ is defined in eq. (\ref{temp}).
	
	\subsubsection{Low Temperature Case}
	By using eqs. (\ref{low}) and (\ref{eq29}), the mutual information in the low temperature limit where $z_h\gg l,\,x$ is given by
	\begin{equation}\label{eq30}
	I_{\text{\tiny low}}=\frac{R^3}{4G_N^{(5)}}\Bigg\{c\left[2\left(\frac{L}{l}\right)^2-\left(\frac{L}{2l+x}\right)^2-\left(\frac{L}{x}\right)^2 \right]
	-\left(\frac{l+x}{z_h}\right)^2\left(\frac{L}{z_h}\right)^2\,f(\xi) \Bigg\}\;,
	\end{equation}
	By eq. (\ref{temp}) we obtain
	\begin{equation}\label{eq30b}
	I_{\text{\tiny low}}=\frac{R^3}{4G_N^{(5)}}\Bigg\{c\left[2\left(\frac{L}{l}\right)^2-\left(\frac{L}{2l+x}\right)^2-\left(\frac{L}{x}\right)^2 \right]
	-\left(\frac{l+x}{l}\right)^2\left(\frac{L}{l}\right)^2\,f(\xi)\,(\pi\hat{T}l)^4\Bigg\}\,,
	\end{equation} 
	where the first terms in brackets matches the result we expect for $T=0$ case \cite{r06} and the finite temperature-dependent term obeys the area-law behavior which has been proved to be true generally in \cite{r05}. 

	\subsubsection{Intermediate Temperature Case}
	In the intermediate temperature limit where $x\ll z_h\ll l\,$, the mutual information is obtained by using eqs. (\ref{eq25}), (\ref{low}) and (\ref{eq29}), and it is given by
	\begin{equation}\label{eq30a}
	I_{\text{\tiny int}}=\frac{R^3}{4G_N^{(5)}}\Bigg\{-c \left(\frac{L}{x}\right)^2+\left(\frac{L}{z_h}\right)^2(\mathcal{S}_4-\frac{\xi}{3})-\left(\frac{x}{z_h}\right)\left(\frac{L}{z_h}\right)^2\sqrt{1+\xi}
 \;- \;\frac{1}{2}\left(\frac{x}{z_h}\right)^2\left(\frac{L}{z_h}\right)^2 f(\xi) \Bigg\},
	\end{equation}
	where $\mathcal{S}_4\equiv \mathcal{S}_1+\mathcal{S}_2+\mathcal{S}_3\,$. As one can see, the mutual information in this limit does not depend on the characteristic length of the system. By using eq. (\ref{temp}) we obtain
	\begin{equation}\label{eq30c}
	I_{\text{\tiny int}}=\frac{R^3L^2}{4G_N^{(5)}}\,(\pi\hat{T})^2\Bigg\{-\frac{c}{(\pi\hat{T}x)^2}+(\mathcal{S}_4-\frac{\xi}{3})-(\pi\hat{T}x)\sqrt{1+\xi}
	-\frac{1}{2}\,f(\xi)\,(\pi\hat{T}x)^2 \Bigg\}\;.
	\end{equation}
	 One could also go further and investigate the case where two strips touch each other i.e. when $x\sim 0\,$. Hence if we take the $x\to 0$ limit of eq. (\ref{eq30c}) we obtain
	\begin{equation}\label{eq30d}
	\lim_{x\to 0}\,I_{\text{\tiny int}}=\frac{R^3}{4G_N^{(5)}}\left\{-c \left(\frac{L}{x}\right)^2+\left(\mathcal{S}_4-\frac{\xi}{3}\right)(\pi\hat{T}L)^2 \right\}\;,
	\end{equation}
	by keeping in mind that in all of the above expressions, $c$ is a numerical coefficient and  $f(\xi)$ depends only on $\xi$ where both are defined in eq. (\ref{eqc}). We note that the leading term in the last expression obeys an area-law divergence with respect to the separation distance $x\,$, and the finite sub-leading term scales with the area of strip, $L^2$, times temperature squared. This area law behavior corresponds to the case where the volume-law  thermodynamic entropy contribution to the entanglement is absent and eq. (\ref{eq30d}) is a measure of pure quantum entanglement. This unique behavior has been also observed for the different backgrounds in \cite{r06,r07}.

	\subsubsection{High Temperature Case}
	As we discussed earlier in this section, for $x/z_h\gg 1$ or $\hat{T}x\gg 1$ we have a vanishing mutual information. It is due to the fact that the minimal surface corresponding to the region $A\cup B$ for large separation distances becomes the disjoint union of the two strips minimal surfaces, hence the mutual information identically vanishes. 

	\section{Mutual Information Near The Critical Point}\label{sec05}
	In this section we study the critical phenomena of the underlying field theory using the information-theoretic measure we introduced in the previous section. Mutual information, a scheme-independent quantity, is considered to serve as an order parameter in the strongly coupled plasma in our setup and we investigate whether the static critical exponents of the theory could be read off from its behavior near or at the critical point\footnote{The role of entanglement entropy as a probe of phase transitions in field theories with holographic dual is pointed out previously in \cite{Nishioka:2006gr,Klebanov:2007ws, Nishioka:2009un}.}. We first begin by recalling the notation we introduced in subsection \ref{sec02} for the critical point which was characterized by the dimensionless quantity $\xi=2\,(1-\sqrt{1-\lambda^2})^2/\lambda^2$ where $\lambda\equiv(\mu/T)/(\pi/\sqrt{2})$. In the critical limit where $\xi\to 2$ or $\lambda\to 1$, we observe that the mutual information, which depends on the parameters of the theory, remains finite and its leading behavior at the critical point, omitting the first constant term in brackets, is proportional to $\sqrt{1-\lambda^2}$ as 
\begin{equation}
I_{\text{\tiny low}}\sim-\frac{R^3}{4G_N^{(5)}}\left(\frac{l+x}{l}\right)^2\left(\frac{L}{l}\right)^2 (\pi\hat{T}l)^4\left((3 b_1+ \frac{2}{3} b_2)-4 (b_1+\frac{2}{3}\,b_2)\sqrt{1-\lambda^2}\right)\;,
\end{equation}	
where we have defined
\begin{equation}\label{b12}
b_1\equiv  \frac{\big(1-w_3+3 w_2+2\,(w_1-1)(\frac{\,a_2}{a_1})\big)}{a_1^2}\;\;\; \text{and}\;\;\;b_2\equiv \frac{\big((w_1-1)(\frac{a_3}{a_1}-1)-8w_2\big)}{a_1^2}\;,	
\end{equation}	
such that $f(\xi)=b_1 (1+\xi)+b_2 \,(\xi^2/6)\,$.	
	It is easy to see that this result, i.e. being proportional to $\sqrt{1-\lambda^2}$, also features in the intermediate regime. Therefore such behavior is independent of the thermal limits and regardless of whether we take the limit where the separation distance $x$ goes to zero or not, it is true for all the results we have obtained so far for the mutual information in subsection \ref{mlimits}. So we can conclude
	 \begin{equation}\label{eq34}
	 I_{\text{\tiny low}}\sim I_{\text{\tiny int}} \propto \left(\frac{\mu}{T}-\frac{\mu_c}{T_c}\right)^{1/2} \;.
	 \end{equation}
	  By comparing eq. (\ref{eq34}) to the expected power-law behavior at the critical point
	  \begin{equation}
	  \left(\frac{\mu}{T}-\frac{\mu_c}{T_c}\right)^{1/\delta}\;,
	  \end{equation}
	  analogous to the power-law behavior of the critical isotherm evaluated at the critical temperature, one may conclude that $\delta=2$\footnote{This result is similar to the critical exponent calculated for this theory using the thermodynamic quantity, charge density.}. Hence by considering the mutual information as an order parameter, we were able to obtain one of the independent critical exponents of the underlying theory. 	
	
In order to obtain the other remaining independent exponent---by following the thermodynamic analogy and the same discussions in the beginning of this section---we can use the slope of mutual information near the critical point for this purpose. We note that although the mutual information is finite there, we see that its derivative with respect to $\lambda$ will tend to infinity as we approach the critical point. For the slope of mutual information in any thermal limit one could write $dI/d\lambda =(dI/d\xi)\,(d\xi/d\lambda)$ where
	\begin{equation}\label{eq32}
	\addtolength{\jot}{1em}
	\frac{d \xi}{d\lambda}=\frac{4 \left(1-\sqrt{1-\lambda^2}\right)^2}{\lambda^3\sqrt{1-\lambda^2}}\,.
	\end{equation}
	Therefore at the critical point, one could easily see that $d\xi/d\lambda$ behaves as $(1-\lambda^2)^{-1/2}$ hence it diverges. The only remaining fact that needs to be checked is whether $d I/d\xi$ is finite or it tends to zero at the critical point. By using eqs. (\ref{eq30}) and (\ref{eq30a}) we obtain
	\begin{equation}\label{eq32b}
	\frac{dI_{\text{\tiny low}}}{d\xi}=-\frac{R^3L^2}{4G_N^{(5)}}
	\frac{\left(l+x\right)^2}{z_h^4}\Big(b_1+\frac{\xi}{3} b_2\Big)\;,
	\end{equation}
	and
	\begin{equation}\label{eq32c}
	\frac{dI_{\text{\tiny int}}}{d\xi}=\frac{R^3L^2}{4G_N^{(5)}} \bigg[\frac{1}{z_h^2}\left(\frac{d\mathcal{S}_4}{d\xi}-\frac{1}{3}\right)-\frac{x}{2 z_h^3\sqrt{1+\xi}}
	-\frac{x^2}{2 z_h^4 a_1^2}\Big(b_1+\frac{\xi}{3} b_2 \Big) \bigg]\;,
	\end{equation}
	as well as
	\begin{equation}\label{eq32d}
		\frac{dI_{\text{\tiny int}}}{d\xi}\Big|_{x\to 0}=\frac{R^3L^2}{4G_N^{(5)}}\bigg[\frac{1}{z_h^2}\left(\frac{d\mathcal{S}_4}{d\xi}-\frac{1}{3}\right) \bigg]\;,
	\end{equation}
	where $b_1$ and $b_2$ are defined in eq. (\ref{b12}). We can see that in all cases, $d I/d\xi$ remains finite at the critical point $\xi=2$. Therefore we reach the conclusion that the mutual information diverges near the critical point with the power-law behavior given by
	\begin{equation}\label{eq33}
	\left(\frac{\mu}{T}-\frac{\mu_c}{T_c}\right)^{-1/2}\equiv \left(\frac{\mu}{T}-\frac{\mu_c}{T_c}\right)^{-\gamma}\;,
	\end{equation}
	 where $\gamma=1/2$ is the critical exponent of this theory identical to the one obtained from the divergence of the R-charge susceptibility defined in eq. (\ref{sus}) near the critical point\footnote{By assuming the correspondence between entanglement entropy and its thermodynamic counterpart and using the same arguments we made in the beginning of this section, we could calculate the slope of entanglement entropy in eqs. (\ref{low}) and (\ref{eq26c}) in order to obtain the exponent $\alpha$ instead, which is analogous to the exponent of specific heat capacity at constant chemical potential, $C_\mu$, evaluated near the critical point. In doing so, we obtain $\alpha=1/2$ which is in full agreement with our results.}. Finally, by using the following known scaling relations for the static critical exponents
	 \begin{equation}
	 \alpha+\beta\left(1+\delta\right)=2 \qquad,\qquad \alpha+2\beta+\gamma=2\;,
	 \end{equation}
	 we obtain $\beta=1/2$ and $\alpha=1/2$\footnote{We could use different names and notations for these critical exponents as these labels are associated with the behavior of quantities in the vicinity of the critical point, approached along the first-order line except for the critical isotherm, while there is no such first-order transition in this model and the phase diagram is one-dimensional. However, to avoid any confusion we would rather use these notations instead.}.
	
	Remarkably, these exponents are identical to those calculated previously for this model within the thermodynamic framework  \cite{Cai:1998ji,Cvetic:1999rb}. The dynamic critical exponent of this model has been also obtained via different quantities in \cite{Finazzo:2016psx, Ebrahim:2018uky, Ebrahim:2017gvk}. It is interesting to note that the same identical values for these four static critical exponents have been also obtained for completely different gravitational backgrounds such as Born-Infeld AdS black holes and topological charged black holes in Horava-Lifshitz gravity \cite{Banerjee:2011cz,Majhi:2012fz,Ma:2014tka}.
	 
\noindent\textbf{Summary}\;
In this work we have argued that information-theoretic measures like mutual information could also be used in order to study the critical phenomena of the strongly coupled field theories in the large-N limit. We based our claim on the result of our analytic calculations for the entanglement entropy and mutual information for the strongly coupled plasma at finite temperature and chemical potential with a critical point using the holographic methods. It is known, as we have also observed here, that despite the volume-law behavior of entanglement entropy in the high temperature limit, mutual information scales with the area of the system, therefore it has the upper hand in capturing the full quantum entanglement structure of the field theories. Based on this observation, we analyzed the critical behavior of the underlying plasma using our analytical results for the mutual information in various thermal limits and we found out that although it was constant at the critical point with the exponent $\delta^{-1}=1/2\,$, it had a power-law divergent slope with the exponent $\gamma=1/2\,$ and therefore we obtained
	  \begin{equation}\label{eq35}
	 \left(\alpha,\beta,\gamma,\delta\right)=\left(\frac{1}{2},\frac{1}{2},\frac{1}{2},2\right)\;,
	 \end{equation}
	 which is in exact agreement with the prior thermodynamics results in the literature. Since entanglement entropy (hence mutual information) has more advantages than the thermodynamic entropy\footnote{Although we should point out that the exact equivalence of entanglement entropy with Bekenstein-Hawking entropy is not clear enough as discussed in  \cite{Solodukhin:2011gn}.}, and it captures the critical phenomena as well, our result suggests that it would be a proper candidate for further investigations regarding the various physical properties of the strongly coupled systems, specially in the ongoing research program of understanding the rich phase structure of hot QCD at finite density.  
	 \section*{Acknowledgment}
	 H. E. would like to thank high energy, cosmology and astroparticle physics group at ICTP where the main parts of the calculations of this paper was done and K. Papadodimas for their warm hospitality. H. E. would also like to thank M. Ali-Akbari for fruitful discussions.  

	\appendix
	\section{Mathematical Relations}\label{app1}
	In this appendix we present some useful relation which we used in our work.
	\begin{itemize}
		\item {\textbf{Newton's binomial and trinomial expansion}}\;\; 
Newton's generalized binomial expansion for $\lvert y\rvert<\lvert x\rvert$  is given by
\begin{equation}
		\begin{aligned}
		\left(x+y\right)^r&=\sum_{k=0}^{\infty} \binom{r}{k}\,x^{r-k}\,y^k\,,\\
		\left(x+y\right)^{-r}&=\sum_{k=0}^{\infty}(-1)^k \binom{r+k-1}{k}\,x^{-r-k}\,y^k\,.
		\end{aligned}
		\end{equation}
		Similarly the generalized trinomial expansion for $\lvert y+z\rvert<\lvert x\rvert$ is given by
		\begin{equation}
		\begin{aligned}
		\left(x+y+z\right)^r&=\sum_{k=0}^{\infty}\sum_{j=0}^{k} \binom{r}{k}\,\binom{k}{j}\,x^{r-k}\,y^{k-j}\,z^j\,,\\
		\left(x+y+z\right)^{-r}&=\sum_{k=0}^{\infty}\sum_{j=0}^{k}(-1)^k \binom{r+k-1}{k}\,\binom{k}{j}\,x^{-r-k}\,y^{k-j}\,z^j\,,
		\end{aligned}
		\end{equation}
		where $x,\,y,\,r\in \mathbb{R}\,$ and $r>0$. Note that for any real numbers $p$ and $q$ we have
		\begin{equation}
		\binom{p}{q}=\frac{\Gamma(p+1)}{\Gamma(q+1)\,\Gamma(p-q+1)}\,.
		\end{equation}
		\item {\textbf{Asymptote of Polylogarithm}}	\;\;
By analytic continuation, polylogarithm function, $\text{Li}_{s}(z)\,$, can be extended to $\lvert z\rvert\ge 1\,$. For $\mathfrak{Re}(s)>0$ and $\lvert z\rvert> 1\,$ its leading term is given by \cite{r11}
		\begin{equation}\label{eq18b}
		\text{Li}_s(z)\sim -\frac{\left[\ln\left(z\right)\right]^s}{\Gamma\left(s+1\right)}\,.
		\end{equation}
	\end{itemize}
	\section{Numerical Constants}\label{app2}
	Here is the list of all numerical constants defined throughout the paper:
	\begin{equation}\label{eq19}
	\begin{aligned}
	a_1&\equiv \sum_{j=0}^{\infty}\frac{\Gamma\left(j+\frac{1}{2}\right)}{\sqrt{\pi}\,\Gamma\left(j+1\right)\,(2+3j)}= \frac{3 \sqrt{\pi }\, \Gamma \left(\frac{5}{3}\right)}{\Gamma \left(\frac{1}{6}\right)}\,,\\
	a_2&\equiv \sum_{j=0}^{\infty}\frac{\Gamma\left(j+\frac{1}{2}\right)}{\sqrt{\pi}\,\Gamma\left(j+1\right)\,(4+3j)}= \frac{\sqrt{\pi }\, \Gamma \left(\frac{7}{3}\right)}{4\, \Gamma \left(\frac{11}{6}\right)}\,,\\
	a_3&\equiv \sum_{j=0}^{\infty}\frac{\Gamma\left(j+\frac{1}{2}\right)\,(4-j)}{\sqrt{\pi }\,\Gamma\left(j+1\right)\,(2+3j)\,(4+3j)}\\
	&= \frac{3}{\sqrt{\pi}} \left[\Gamma \left(\frac{5}{6}\right) \Gamma \left(\frac{5}{3}\right)-\frac{3}{5}\,\Gamma \left(\frac{7}{6}\right) \Gamma
	\left(\frac{7}{3}\right)\right]-\frac{1}{70} \, _3F_2\left(\frac{3}{2},\frac{5}{3},\frac{7}{3};\frac{8}{3},\frac{10}{3};1\right)\,,
\end{aligned}
	\end{equation}
and
\begin{equation}\label{w}
	\begin{aligned}	
w_1&\equiv \frac{1}{\sqrt{\pi}}\sum_{j=1}^{\infty}\frac{\Gamma\left(j+\frac{1}{2}\right)}{\Gamma\left(j+1\right)\left(3j-1\right)}= \frac{1}{2^{2/3}}{\, _2F_1\left(\frac{1}{3},\frac{2}{3};\frac{5}{3};-1\right)}\,,\\
	w_2&\equiv \frac{1}{\sqrt{\pi}}\sum_{j=1}^{\infty}\frac{j\, \Gamma\left(j+\frac{1}{2}\right)}{ \Gamma\left(j+1\right)\,\left(3j-1\right)\,(3j+1)}=\frac{1}{16} \, _3F_2\left(\frac{2}{3},\frac{4}{3},\frac{3}{2};\frac{5}{3},\frac{7}{3};1\right)\,,\\
	w_3&\equiv \frac{1}{\sqrt{\pi}}\sum_{j=1}^{\infty}\frac{\Gamma\left(j+\frac{1}{2}\right)}{ \Gamma\left(j+1\right)\,\left(3j-1\right)\,(3j+1)}\\
	&=\frac{3}{16} \, _3F_2\left(\frac{2}{3},\frac{4}{3},\frac{3}{2};\frac{5}{3},\frac{7}{3};1\right)-\frac{1}{\sqrt[3]{2}}\,_2F_1\left(\frac{4}{3},\frac{5}{3};\frac{7}{3};-1\right)\,.
	\end{aligned}
	\end{equation}
	\vfill
	\section{Minimal Surface Area in the High Temperature Limit}\label{app3}
	The regularized area of eq. (\ref{eq21}) in the high temperature limit is given by
	\begin{equation}\label{eq19b}
	\begin{aligned}
	\mathcal{A}^{\text{\tiny finite}}_{\text{\tiny high}}&=\frac{L^2R^3l}{z_c^3}\left(1+\xi\left(\frac{z_c}{z_h}\right)^2\right)^{\frac{1}{2}}\\
	&+\frac{L^2R^3}{z_c^2}\Biggl\{\frac{3\,\xi}{2}\left(\frac{z_c}{z_h}\right)^2-\frac{4}{5}\left[1+\xi\left(\frac{z_c}{z_h}\right)^2\right]^{\frac{3}{2}}+\frac{1+\xi}{3\xi }\left(\frac{z_c}{z_h}\right)^2\left[\left(1+\xi\left(\frac{z_c}{z_h}\right)^2\right)^{\frac{3}{2}}-1\right]\\
	&\qquad\qquad-\frac{28}{15\,\xi^2}\left(\frac{z_h}{z_c}\right)^4 \left[1+\xi\left(\frac{z_c}{z_h}\right)^2\right]^{\frac{3}{2}}\left[\frac{\xi}{2}\left(\frac{z_c}{z_h}\right)^2+\left[1+\xi\left(\frac{z_c}{z_h}\right)^2\right]^{-\frac{1}{2}}-1\right]\\
	&\qquad\qquad-\frac{1+\xi}{35\,\xi^4}\left(\frac{z_h}{z_c}\right)^4 \left[1+\xi\left(\frac{z_c}{z_h}\right)^2\right]^{\frac{3}{2}}\Biggl[-16+\frac{16}{\sqrt{1+\xi\left(\frac{z_c}{z_h}\right)^2}}+\xi\left(\frac{z_c}{z_h}\right)^2\\
	&\qquad\qquad\qquad\qquad\qquad\qquad\qquad\qquad\qquad\,\,\,\times\left(8-6\,\xi\left(\frac{z_c}{z_h}\right)^2+5\xi^2\left(\frac{z_c}{z_h}\right)^4\right) \Biggr] \Biggr\}\\
	&+\frac{L^2R^3}{z_c^2}\Biggl\{\sum_{k=2}^{\infty}\sum_{n=0}^{k}\sum_{m=0}^{\infty}\frac{\Gamma\left(k+\frac{1}{2}\right)\,\Gamma\left(m+\frac{1}{2}\right)\,\Gamma\left(k+n+2\right)\,}{\pi \,\Gamma\left(n+1\right)\,\Gamma\left(k-n+1\right)\,\Gamma\left(k+n+m+3\right)\,}\\
	&\qquad\qquad\times (-1)^{k+n}\,\xi^{k-n+m}\,(1+\xi)^{n}\left[1+\xi\left(\frac{z_c}{z_h}\right)^2\right]^{-m-\frac{1}{2}}\left(\frac{z_c}{z_h}\right)^{2(k+n+m)}\\
	&\qquad\qquad\times \Biggl\{\frac{m+1}{k+n-1}\left[1+\frac{m+1}{k+n}\left(2+\frac{m}{k+n+1}\right)\right]\\
	&\qquad\qquad\qquad+\frac{\left(1+\xi\left(\frac{z_c}{z_h}\right)^2\right)(m+1)}{k+n}\left(2+\frac{m}{k+n+1}\right) \Biggr\}\Biggr\}\\
	&+\frac{L^2R^3}{z_c^2}\Biggl\{\sum_{k=2}^{\infty}\sum_{n=0}^{k}\sum_{m=0}^{\infty}\sum_{j=1}^{\infty}\frac{\Gamma\left(k+\frac{1}{2}\right)\,\Gamma\left(j+m+\frac{1}{2}\right)\,\Gamma\left(k+n+3j+2\right)\,}{\pi\,\Gamma\left(j+1\right)\, \,\Gamma\left(n+1\right)\,\Gamma\left(k-n+1\right)\,\Gamma\left(k+n+m+3j+3\right)\,}\\
	&\qquad\qquad\times (-1)^{k+n}\,\xi^{k-n+m}\,(1+\xi)^{n}\left[1+\xi\left(\frac{z_c}{z_h}\right)^2\right]^{-m-\frac{1}{2}}\left(\frac{z_c}{z_h}\right)^{2(k+n+m)}\\
	&\qquad\qquad\times \Biggl\{\frac{m+1}{k+n+3j-1}\left[1+\frac{m+1}{k+n+3j}\left(2+\frac{m}{k+n+3j+1}\right)\right]\\
	&\qquad\qquad\qquad+\frac{\left(1+\xi\left(\frac{z_c}{z_h}\right)^2\right)(m+1)}{k+n+3j}\left(2+\frac{m}{k+n+3j+1}\right) \Biggr\}\Biggr\}\,.
	\end{aligned}
	\end{equation}
	\section{Sub-leading Corrections in the Near Horizon Limit}\label{app4}
	In this appendix we will investigate the convergence of characteristic length and behavior of area for $z_c\to z_h\,$. We note that the large terms of the series in eq. (\ref{eq15}) for the characteristic length scale $l$ grow as\footnote{By approximating the series in the limit where all the free indices are set to infinity.}
	\begin{equation}\label{eq19c}
	\quad 3^{-m}\,  \xi ^{m}\, k^{-1/2} \left(1+\xi\right)^k\,j^{-3/2}\left[1+\xi\left(\frac{z_c}{z_h}\right)^2\right]^{-m}\left(\frac{z_c}{z_h}\right)^{2
		(2k+m)}\;,
	\end{equation}
	which diverges for $z_c=z_h\,$. We can overcome this situation by isolating the divergent term of eq. (\ref{eq19c}) from eq. (\ref{eq15}) so that $l$ converges. Hence the regularized $l$ becomes
	\begin{equation}\label{eq19d}
	\begin{aligned}
	\frac{l}{2}&=z_c\,\sum_{k=1}^{\infty}\sum_{m=1}^{\infty}\sum_{j=1}^{\infty}\Biggl\{ \sum_{n=1}^{k}\Bigg\{  G_{knmj}\,F_{knmj} \left(\frac{z_c}{z_h}\right)^{2
		(k+m+n)}\Bigg \}	\\
	&\qquad\qquad\qquad\qquad-\frac{3^{-m}\,  \xi ^{m}\,\left(1+\xi\right)^k}{6\pi \, k^{1/2}\,j^{3/2}}\left[1+\xi\left(\frac{z_c}{z_h}\right)^2\right]^{-m}\left(\frac{z_c}{z_h}\right)^{2
		(2k+m)} \Biggr\}\\
	&+\frac{z_c}{4}+\frac{z_c\,\xi}{6\pi}\left[3+2\,\xi\left(\frac{z_c}{z_h}\right)^2 \right]^{-1}\left(\frac{z_c}{z_h}\right)^2\zeta\left(\frac{3}{2}\right)\text{Li}_{\frac{1}{2}}\left(\left(1+\xi\right)\left(\frac{z_c}{z_h}\right)^4\right)\;.
	\end{aligned}
	\end{equation}
	where we made use of the following relations for eq. (\ref{eq19c}) in the process of regularization
	\begin{equation}\label{zeta}
	\begin{aligned}
	\text{Li}_s(z)&=\sum_{k=1}^{\infty}\frac{z^k}{k^s}\;,\\
	\zeta(p)&=\sum_{j=1}^{\infty}\frac{1}{j^p}\;,\\
	\end{aligned}
	\end{equation}
	and the fact that remaining summation over $m$ in eq. (\ref{eq19c}) can be performed. Note that in eq. (\ref{zeta}), $\zeta(p)$ is the Riemann zeta function and $\text{Li}_s(z)$ is the polylogarithm function of order $s\,$. As a mathematical curiosity, one might consider the appearance of Riemann zeta function and polylogarithm with rational order (or even with the integer order)\footnote{The case for $s\in \mathbb{N}$ in both $\zeta(s)$ and $\text{Li}_s(z)$ is the subject of wide interest in the number theory literature. See for example \cite{r12}.} an interesting phenomena due to their direct link to number theory.
	
	Now since the minimal surface remains at a finite distance from horizon \cite{r13}, we could safely assume $z_c=z_h(1-\varepsilon)\,$, where $\varepsilon< 1\,$. Then by the help of eq. (\ref{eq18b}), i.e. expanding the polylogarithm in eq. (\ref{eq19d}) and then solving the result for $\varepsilon$ at leading order, we obtain
	\begin{equation}\label{eq19e}
	\varepsilon=\frac{1}{2}\ln\left(1+\xi\right)-\frac{3\pi^{3/2}\left(3+2\,\xi \right)\sqrt{\ln\left(1+\xi\right)}}{4\,\zeta\left(\frac{3}{2} \right)\,\xi}\left[\sigma_1+\frac{1}{2}-\left(\frac{l}{z_h}\right)\right]\;,
	\end{equation}
	where we defined
	\begin{equation}
	\begin{aligned}
	\sigma_1&\equiv \sum_{k=1}^{\infty}\sum_{n=1}^{k}\sum_{\substack{j=1\\m=1}}^{\infty} \frac{(-1)^{k+n}\, \Gamma \left(k+\frac{1}{2}\right) \Gamma \left(j+m+\frac{1}{2}\right) (1+\xi )^{n-m}\, \Gamma (3 j+k+n+2)\, \xi ^{k+m-n}}{\pi\,\Gamma (j+1) \Gamma
		(n+1) \Gamma (k-n+1) \Gamma (3 j+k+m+n+3)} \\
	&-\sum_{k=1}^{\infty}\sum_{m=1}^{\infty}\sum_{j=1}^{\infty}\frac{3^{-m-1}\,  \xi ^{m}\,\left(1+\xi\right)^{k-m}}{\pi \, k^{1/2}\,j^{3/2}}\;.
	\end{aligned}
	\end{equation}
	Finally, we are ready to calculate the sub-leading corrections to the minimal surface area in the near horizon limit. Similarly, we observe that the large terms of the series in eq. (\ref{eq19b}) for the area behave as
	\begin{equation}
	\quad 3^{-m}\,\xi^m\,(1+m)\,\left(1+\xi\right)^{k}\,k^{-1/2}\,j^{-5/2}\left[1+\xi\left(\frac{z_c}{z_h}\right)^2\right]^{-m}\left(\frac{z_c}{z_h}\right)^{2
		(2k+m)}\;.
	\end{equation}
	Hence by following the same regularization procedure as we did for $l$ by isolating this piece from eq. (\ref{eq19b}) and performing its sum, together with the assumption  $z_c=z_h(1-\varepsilon)\,$, we expand the resulted expression at the first order in $\varepsilon$ and by the help of eq. (\ref{eq19e}) we obtain 
	\begin{equation}\label{eq20a}
	\begin{aligned}
	\mathcal{A}^{\text{\tiny finite}}_{\text{\tiny high}}&=R^3\left(\frac{L}{z_h}\right)^2\left(\frac{l}{z_h}\right)\sqrt{1+\xi}+R^3\left(\frac{L}{z_h}\right)^2\left(\mathcal{S}_1+\mathcal{S}_2+\mathcal{S}_3 \right)\\
	&-R^3\left(\frac{L}{z_h}\right)^2\Biggl\{ \frac{3\left(1+\xi\right)^{3/2}\zeta\left(\frac{5}{2} \right)}{\zeta\left(\frac{3}{2} \right)\xi}\left[\sigma_1+\frac{1}{2}-\left(\frac{l}{z_h}\right) \right]+\frac{\left(1+\xi\right)^{5/2}\zeta\left(\frac{5}{2} \right)}{\pi\left(3+2\,\xi\right)}+\sigma_2
	\Biggr\}\;,
	\end{aligned}
	\end{equation}
	where 
	\begin{equation}
	\sigma_2\equiv  \sum_{k=2}^{\infty}\sum_{j=1}^{\infty}\sum_{m=0}^{\infty}\frac{3^{-m}\,(1+m)\left(3+2\,\xi\right)\,\xi^m\,\left(1+\xi\right)^{k-m}}{9\pi\,\sqrt{1+\xi}\,k^{1/2}\,j^{5/2}}\;.
	\end{equation}
	Note that the expression in the second line of eq. (\ref{eq20a}) is the desired sub-leading contribution to the area of the minimal surface in the high temperature limit.
	

\begin{thebibliography}{99}
		\bibitem{Rangamani:2016dms}
		M.~Rangamani and T.~Takayanagi,
		``Holographic Entanglement Entropy,''
		Lect.\ Notes Phys.\  {\bf 931}, pp.1 (2017)
		\href{https://arxiv.org/abs/1609.01287}{[hep-th/1609.01287]}.
		\bibitem{Holzhey:1994we} 
		C.~Holzhey, F.~Larsen and F.~Wilczek,
		``Geometric and renormalized entropy in conformal field theory,''
		Nucl.\ Phys.\ B {\bf 424}, 443 (1994)
		\href{https://arxiv.org/abs/9403108}{[hep-th/9403108]}.
		
		\bibitem{Calabrese:2004eu} 
		P.~Calabrese and J.~L.~Cardy,
		``Entanglement entropy and quantum field theory,''
		J.\ Stat.\ Mech.\  {\bf 0406}, P06002 (2004)
		\href{https://arxiv.org/abs/0405152}{[hep-th/0405152]}.
		\bibitem{Casini:2009sr} 
		H.~Casini and M.~Huerta,
		``Entanglement entropy in free quantum field theory,''
		J.\ Phys.\ A {\bf 42}, 504007 (2009)
		\href{https://arxiv.org/abs/0905.2562}{[hep-th/0905.2562]}.
		
		\bibitem{Nishioka:2018khk} 
		T.~Nishioka,
		``Entanglement entropy: holography and renormalization group,''
		Rev.\ Mod.\ Phys.\  {\bf 90}, no. 3, 035007 (2018)
		\href{https://arxiv.org/abs/1801.10352}{[hep-th/1801.10352]}.
		
		\bibitem{Bombelli:1986rw} 
		L.~Bombelli, R.~K.~Koul, J.~Lee and R.~D.~Sorkin,
		``A Quantum Source of Entropy for Black Holes,''
		Phys.\ Rev.\ D {\bf 34}, 373 (1986).
		
		\bibitem{Srednicki:1993im}
		M.~Srednicki,
		``Entropy and area,''
		Phys.\ Rev.\ Lett.\  {\bf 71}, 666 (1993)
		\href{https://arxiv.org/abs/hep-th/9303048}{[hep-th/9303048]}.
		
		\bibitem{Maldacena:1997re} 
		J.~M.~Maldacena,
		``The Large N limit of superconformal field theories and supergravity,''
		Int.\ J.\ Theor.\ Phys.\  {\bf 38}, 1113 (1999)
		[Adv.\ Theor.\ Math.\ Phys.\  {\bf 2}, 231 (1998)]
		\href{https://arxiv.org/abs/hep-th/9711200}{[hep-th/9711200]}.
		
		\bibitem{Gubser:1998bc} 
		S.~S.~Gubser, I.~R.~Klebanov and A.~M.~Polyakov,
		``Gauge theory correlators from noncritical string theory,''
		Phys.\ Lett.\ B {\bf 428}, 105 (1998)
		\href{https://arxiv.org/abs/hep-th/9802109}{[hep-th/9802109]}.
		
		\bibitem{Witten:1998qj} 
		E.~Witten,
		``Anti-de Sitter space and holography,''
		Adv.\ Theor.\ Math.\ Phys.\  {\bf 2}, 253 (1998)
		\href{https://arxiv.org/abs/hep-th/9802150}{[hep-th/9802150]}.
		
		\bibitem{r01}
		S.~Ryu and T.~Takayanagi,
		``Holographic derivation of entanglement entropy from AdS/CFT,''
		Phys.\ Rev.\ Lett.\  {\bf 96}, 181602 (2006)
		\href{https://arxiv.org/abs/hep-th/0603001}{[hep-th/0603001]}.
		
		\bibitem{r02}
		V.~E.~Hubeny, M.~Rangamani and T.~Takayanagi,
		``A Covariant holographic entanglement entropy proposal,''
		JHEP {\bf 0707}, 062 (2007)
		\href{https://arxiv.org/abs/0705.0016}{[hep-th/0705.0016]}.
		
		
		\bibitem{Headrick:2007km} 
		M.~Headrick and T.~Takayanagi,
		``A Holographic proof of the strong subadditivity of entanglement entropy,''
		Phys.\ Rev.\ D {\bf 76}, 106013 (2007)
		\href{https://arxiv.org/abs/0704.3719}{[hep-th/0704.3719]}.
		
		\bibitem{r09}
		A.~C.~Wall,
		``Maximin Surfaces, and the Strong Subadditivity of the Covariant Holographic Entanglement Entropy,''
		Class.\ Quant.\ Grav.\  {\bf 31}, no. 22, 225007 (2014)
		\href{https://arxiv.org/abs/1211.3494}{[hep-th/1211.3494]}.
		
		\bibitem{Ryu:2006ef} 
		S.~Ryu and T.~Takayanagi,
		``Aspects of Holographic Entanglement Entropy,''
		JHEP {\bf 0608}, 045 (2006)
		\href{https://arxiv.org/abs/hep-th/0605073}{[hep-th/0605073]}.
		
		\bibitem{Casini:2011kv} 
		H.~Casini, M.~Huerta and R.~C.~Myers,
		``Towards a derivation of holographic entanglement entropy,''
		JHEP {\bf 1105}, 036 (2011)
		\href{https://arxiv.org/abs/1102.0440}{[hep-th/1102.0440]}.
		
		\bibitem{Rahimi:2016bbv} 
		M.~Rahimi, M.~Ali-Akbari and M.~Lezgi,
		``Entanglement entropy in a non-conformal background,''
		Phys.\ Lett.\ B {\bf 771}, 583 (2017)
		\href{https://arxiv.org/abs/1610.01835}{[hep-th/1610.01835]}.
		
		\bibitem{Lokhande:2017jik} 
		S.~F.~Lokhande, G.~W.~J.~Oling and J.~F.~Pedraza,
		``Linear response of entanglement entropy from holography,''
		JHEP {\bf 1710}, 104 (2017)
		\href{https://arxiv.org/abs/1705.10324}{[hep-th/1705.10324]}.
		
		\bibitem{Myers:2012ed} 
		R.~C.~Myers and A.~Singh,
		``Comments on Holographic Entanglement Entropy and RG Flows,''
		JHEP {\bf 1204}, 122 (2012)
		\href{https://arxiv.org/abs/1202.2068}{[hep-th/1202.2068]}.
		
		\bibitem{Asadi:2018ijf} 
		M.~Asadi and M.~Ali-Akbari,
		``Holographic Mutual and Tripartite Information in a Symmetry Breaking Quench,''
		Phys.\ Lett.\ B {\bf 785}, 409 (2018)
		\href{https://arxiv.org/abs/1804.05604}{[hep-th/1804.05604]}.
		
		\bibitem{BabaeiVelni:2019pkw} 
		K.~Babaei Velni, M.~R.~Mohammadi Mozaffar and M.~H.~Vahidinia,
		``Some Aspects of Entanglement Wedge Cross-Section,''
		JHEP {\bf 1905}, 200 (2019)
		\href{https://arxiv.org/abs/1903.08490}{[hep-th/1903.08490]}.
		
		\bibitem{Fareghbal:2019czx} 
		R.~Fareghbal and M.~Hakami Shalamzari,
		``First Law of Entanglement Entropy in Flat-Space Holography,''
		Phys.\ Rev.\ D {\bf 100}, no. 10, 106006 (2019)
		\href{https://arxiv.org/abs/1908.02560}{[hep-th/1908.02560]}.
		
		\bibitem{r04}
		H.~Casini and M.~Huerta,
		``A Finite entanglement entropy and the c-theorem,''
		Phys.\ Lett.\ B {\bf 600}, 142 (2004)
		\href{https://arxiv.org/abs/hep-th/0405111}{[hep-th/0405111]}.
		
		\bibitem{r05}
		M.~M.~Wolf, F.~Verstraete, M.~B.~Hastings and J.~I.~Cirac,
		``Area Laws in Quantum Systems: Mutual Information and Correlations,''
		Phys.\ Rev.\ Lett.\  {\bf 100}, no. 7, 070502 (2008)
		\href{https://arxiv.org/abs/0704.3906}{[quant-ph/0704.3906]}.
		
		\bibitem{Headrick:2010zt} 
		M.~Headrick,
		``Entanglement Renyi entropies in holographic theories,''
		Phys.\ Rev.\ D {\bf 82}, 126010 (2010)
		\href{https://arxiv.org/abs/1006.0047}{[hep-th/1006.0047]}.
		
		\bibitem{Gubser:1998jb} 
		S.~S.~Gubser,
		``Thermodynamics of spinning D3-branes,''
		Nucl.\ Phys.\ B {\bf 551}, 667 (1999)
		\href{https://arxiv.org/abs/hep-th/9810225}{[hep-th/9810225]}.
		
		\bibitem{Behrndt:1998jd} 
		K.~Behrndt, M.~Cvetic and W.~A.~Sabra,
		``Nonextreme black holes of five-dimensional N=2 AdS supergravity,''
		Nucl.\ Phys.\ B {\bf 553}, 317 (1999)
		\href{https://arxiv.org/abs/hep-th/9810227}{[hep-th/9810227]}.
		
		\bibitem{Kraus:1998hv} 
		P.~Kraus, F.~Larsen and S.~P.~Trivedi,
		``The Coulomb branch of gauge theory from rotating branes,''
		JHEP {\bf 9903}, 003 (1999)
		\href{https://arxiv.org/abs/hep-th/9811120}{[hep-th/9811120]}.
		
		\bibitem{Cvetic:1999ne} 
		M.~Cvetic and S.~S.~Gubser,
		``Phases of R charged black holes, spinning branes and strongly coupled gauge theories,''
		JHEP {\bf 9904}, 024 (1999)
		\href{https://arxiv.org/abs/hep-th/9902195}{[hep-th/9902195]}.
		
		\bibitem{Cvetic:1999rb} 
		M.~Cvetic and S.~S.~Gubser,
		``Thermodynamic stability and phases of general spinning branes,''
		JHEP {\bf 9907}, 010 (1999)
		\href{https://arxiv.org/abs/hep-th/9903132}{[hep-th/9903132]}.
		
		\bibitem{DeWolfe:2010he} 
		O.~DeWolfe, S.~S.~Gubser and C.~Rosen,
		``A holographic critical point,''
		Phys.\ Rev.\ D {\bf 83}, 086005 (2011)
		\href{https://arxiv.org/abs/1012.1864}{[hep-th/1012.1864]}.
		
		\bibitem{r00a}
		O.~DeWolfe, S.~S.~Gubser and C.~Rosen,
		``Dynamic critical phenomena at a holographic critical point,''
		Phys.\ Rev.\ D {\bf 84}, 126014 (2011)
		\href{https://arxiv.org/abs/1108.2029}{[hep-th/1108.2029]}.
		
		\bibitem{Cai:1998ji} 
		R.~G.~Cai and K.~S.~Soh,
		``Critical behavior in the rotating D-branes,''
		Mod.\ Phys.\ Lett.\ A {\bf 14}, 1895 (1999)
		\href{https://arxiv.org/abs/hep-th/9812121}{[hep-th/9812121]}.
		
		
		\bibitem{r08}
		P.~Hayden, M.~Headrick and A.~Maloney,
		``Holographic Mutual Information is Monogamous,''
		Phys.\ Rev.\ D {\bf 87}, no. 4, 046003 (2013)
		\href{https://arxiv.org/abs/1107.2940}{[hep-th/1107.2940]}.
		
		\bibitem{Fischler:2012ca} 
		W.~Fischler and S.~Kundu,
		``Strongly Coupled Gauge Theories: High and Low Temperature Behavior of Non-local Observables,''
		JHEP {\bf 1305}, 098 (2013)
		\href{https://arxiv.org/abs/1212.2643}{[hep-th/1212.2643]}.
		
		
		\bibitem{r06}
		W.~Fischler, A.~Kundu and S.~Kundu,
		``Holographic Mutual Information at Finite Temperature,''
		Phys.\ Rev.\ D {\bf 87}, no. 12, 126012 (2013)
		\href{https://arxiv.org/abs/1212.4764}{[hep-th/1212.4764]}.
		
		\bibitem{r07}
		S.~Kundu and J.~F.~Pedraza,
		``Aspects of Holographic Entanglement at Finite Temperature and Chemical Potential,''
		JHEP {\bf 1608}, 177 (2016)
		\href{https://arxiv.org/abs/1602.07353}{[hep-th/1602.07353]}.
		
		\bibitem{r13}
		V.~E.~Hubeny,
		``Extremal surfaces as bulk probes in AdS/CFT,''
		JHEP {\bf 1207}, 093 (2012)
		\href{https://arxiv.org/abs/1203.1044}{[hep-th/1203.1044]}.
		
		\bibitem{Nishioka:2006gr} 
		T.~Nishioka and T.~Takayanagi,
		``AdS Bubbles, Entropy and Closed String Tachyons,''
		JHEP {\bf 0701}, 090 (2007)
		\href{https://arxiv.org/abs/hep-th/0611035}{[hep-th/0611035]}.
		
		\bibitem{Klebanov:2007ws} 
		I.~R.~Klebanov, D.~Kutasov and A.~Murugan,
		``Entanglement as a probe of confinement,''
		Nucl.\ Phys.\ B {\bf 796}, 274 (2008)
		\href{https://arxiv.org/abs/0709.2140}{[hep-th/0709.2140]}.
		
		\bibitem{Nishioka:2009un} 
		T.~Nishioka, S.~Ryu and T.~Takayanagi,
		``Holographic Entanglement Entropy: An Overview,''
		J.\ Phys.\ A {\bf 42}, 504008 (2009)
		\href{https://arxiv.org/abs/0905.0932}{[hep-th/0905.0932]}.
		
		\bibitem{Finazzo:2016psx} 
		S.~I.~Finazzo, R.~Rougemont, M.~Zaniboni, R.~Critelli and J.~Noronha,
		``Critical behavior of non-hydrodynamic quasinormal modes in a strongly coupled plasma,''
		JHEP {\bf 1701}, 137 (2017)
		\href{https://arxiv.org/abs/1610.01519}{[hep-th/1610.01519]}.
		
		\bibitem{Ebrahim:2018uky} 
		H.~Ebrahim, M.~Asadi and M.~Ali-Akbari,
		``Evolution of Holographic Complexity Near Critical Point,''
		JHEP {\bf 1909}, 023 (2019)
		\href{https://arxiv.org/abs/1811.12002}{[hep-th/1811.12002]}.
		
		\bibitem{Ebrahim:2017gvk} 
		H.~Ebrahim and M.~Ali-Akbari,
		``Dynamically probing strongly-coupled field theories with critical point,''
		Phys.\ Lett.\ B {\bf 783}, 43 (2018)
		\href{https://arxiv.org/abs/1712.08777}{[hep-th/1712.08777]}.
		
		\bibitem{Banerjee:2011cz} 
		R.~Banerjee and D.~Roychowdhury,
		``Critical phenomena in Born-Infeld AdS black holes,''
		Phys.\ Rev.\ D {\bf 85}, 044040 (2012)
		\href{https://arxiv.org/abs/1111.0147}{[gr-qc/1111.0147]}.
		
		\bibitem{Majhi:2012fz} 
		B.~R.~Majhi and D.~Roychowdhury,
		``Phase transition and scaling behavior of topological charged black holes in Horava-Lifshitz gravity,''
		Class.\ Quant.\ Grav.\  {\bf 29}, 245012 (2012)
		\href{https://arxiv.org/abs/1205.0146}{[gr-qc/1205.0146]}.
		
		\bibitem{Ma:2014tka} 
		M.~S.~Ma, F.~Liu and R.~Zhao,
		``Continuous phase transition and critical behaviors of 3D black hole with torsion,''
		Class.\ Quant.\ Grav.\  {\bf 31}, 095001 (2014)
		\href{https://arxiv.org/abs/1403.0449}{[gr-qc/1403.0449]}.
		
		\bibitem{Solodukhin:2011gn} 
		S.~N.~Solodukhin,
		``Entanglement entropy of black holes,''
		Living Rev.\ Rel.\  {\bf 14}, 8 (2011)
		\href{https://arxiv.org/abs/1104.3712}{[hep-th/1104.3712]}.
		
		\bibitem{r11}
		D. C. Wood, \emph{"The Computation of Polylogarithms,"} \emph{Technical Report 15-92*}, University of Kent, Canterbury, UK (1992), pg. 182-196 
		
		
		
		\bibitem{r12}
		D. Zagier, \emph{"Values of Zeta Functions and Their Applications,"} First European Congress of Mathematics, Volume II, Progress in Math. {\bf 120}, Birkh{\"a}user-Verlag, Basel, (1994) 497-512
				
	
	\end{thebibliography}
\end{document}